\documentclass[aps,pra,twocolumn,showpacs,superscriptaddress]{revtex4}
\usepackage[T2A]{fontenc}
\usepackage[cp1251]{inputenc}
\usepackage[english]{babel}
\usepackage{graphicx}
\usepackage{color}
\usepackage{bm}
\usepackage{amssymb}
\usepackage[intlimits]{amsmath}
\usepackage{amsmath}

\newcommand{\be}{\begin{equation}}
\newcommand{\ee}{\end{equation}}
\newcommand{\ben}{\begin{equation*}}
\newcommand{\een}{\end{equation*}}

\begin{document}

%\begin{titlepage}
\hskip 2cm
\date{\today}
\title{Mutual information in nonlinear communication channel.  Preliminary analytical results in large SNR and small nonlinearity limit}

\author{I.~S.~Terekhov}
\email[E-mail: ]{I.S.Terekhov@gmail.com} \affiliation{Budker Institute of  Nuclear
Physics of Siberian Branch Russian  Academy of Sciences, Novosibirsk, 630090 Russia}
%\affiliation{School of Physics, University of New South Wales, Sydney 2052,
%Australia}
\affiliation{Novosibirsk State University, Novosibirsk, 630090 Russia}
\author{A.~V.~Reznichenko}
\email[E-mail: ]{A.V.Reznichenko@inp.nsk.su} \affiliation{Budker Institute of
Nuclear Physics of Siberian Branch Russian Academy of Sciences, Novosibirsk, 630090
Russia} \affiliation{Novosibirsk State University, Novosibirsk, 630090 Russia}
\author{S.~K.~Turitsyn}
\email[E-mail: ]{s.k.turitsyn@aston.ac.uk} \affiliation{Novosibirsk State
University, Novosibirsk, 630090 Russia}
\affiliation{Aston Institute of Photonics
Technologies, Aston University, Aston Triangle, Birmingham, B4 7ET, UK}

\begin{abstract}

Applying perturbation theory to the path-integral representation for the mutual
information of the nonlinear communication channel described by the nonlinear
Shr\"{o}dinger equation (NLSE) with the additive Gaussian noise we analyze the
analytical expression for the mutual information at large signal-to-noise ratio
($\mathrm{SNR}$) and small nonlinearity. We classify all possible corrections to the
mutual information in nonlinearity parameter and demonstrate that all singular in
$\mathrm{SNR}$ terms vanish in the final result. Furthermore our analytical result
demonstrates that the corrections to Shannon's contribution to the mutual
information in the leading order in $\mathrm{SNR}$ are of order of squared
nonlinearity parameter. We outline the way for the calculation of these corrections
in the further investigations.

\end{abstract}

\keywords{information entropy, channel capacity, mutual information, nonlinear Shr\"{o}dinger equation.}
\pacs{05.10.Gg, 89.70.-a,  02.70.-c,02.70.Rr,05.90.+m}
\maketitle
%==================Introduction

\section{Introduction}

The channel capacity is one of the central concepts of information theory that has
its roots in statistical physics \cite{Shannon:1948}. The capacity  introduced by
Shannon in his seminal work \cite{Shannon:1948}  gives the maximum rate at which
information can be reliably transmitted through a noisy communication channel. The
channel capacity (in bits per symbol) is formally defined as a maximum of the mutual
information $I_{P[X]}$ over the input signal probability distribution functional
(PDF) $P[X]$:
\begin{eqnarray}
C=\max_{P[X]}  I_{P[X]}, \label{capacity1}
\end{eqnarray} under the condition of a fixed average power. The mutual information
$I_{P[X]}$  (in continuous-input, and continuous-output channel) is expressed
through the path-integral over input $X$ and output $Y$ signals:
\begin{eqnarray}
\begin{split} % I deleted \lim_{T\to\infty}\frac{1}{T}
&I_{P[X]}= \int {\cal D}X {\cal D}Y   P[X] P[Y| X]  \log\Big[ \frac{P[Y|
X]}{P_{out}[Y]}\Big], \label{capacity2}
\end{split}
\end{eqnarray}
where the output signal PDF $P_{out}[Y]$ reads
\begin{align}
& P_{out}[Y]=\int {\cal D}X P[X] P[Y|X],
\label{Pout}
\end{align}
with  $P[Y| X]$ being  the conditional probability density,   that is, the probability
of receiving  output signal $Y$ when the input signal is $X$. Both $X$ and $Y$ may be discrete or continuous. When $X$ is discrete,
notation integral over $X$ stands for the summation of an under integral function over its discrete support. Capacity in bits per symbol multiplied by the rate at which symbols are transmitted (in symbols per second) gives the error-free information transmission rate in bits per second.

The above definition (\ref{capacity2}) exemplifies that  channel capacity has a
close link to information entropy \cite{Shannon:1948}. Mutual information is a
difference between the entropy of the output signal $H[Y]=-\int{\cal D}Y P_{out}[Y]
\log\Big[P_{out}[Y]\Big] $ and conditional entropy $H[Y|X]=-\int {\cal D}X {\cal D}Y
P[X]P[Y| X] \log\Big[P[Y| X]\Big]$ (having a meaning of a measure of the uncertainty
about the output field $Y$ if the input field $X$ is known). When the signal and the
noise are independent variables and the received signal $Y$ is the sum of the
transmitted signal $X$ and the noise, then it can be shown explicitly that the
entropy is generated during transmission in noisy channel: $H[Y] \geq H[X]$. In this
case, the transmission rate is the entropy of the received signal less the entropy
of the noise. The maximum  of the functional (\ref{capacity1}), i.e. the channel
capacity, can be calculated for such linear channels with an additive white Gaussian
noise (AWGN):
\begin{eqnarray}\label{CapacityShannon}
C\propto\log\left(1+\mathrm{SNR}\right)\,,
\end{eqnarray}
where $\mathrm{SNR}$ is a signal-to-noise power ratio \cite{Shannon:1948}.

This seminal theoretical result is the foundation  of the communication theory and
it has proven its importance in a number of practical applications. To some extent,
the Eq. (\ref{CapacityShannon}) worked so well in so many situations that some
engineers cease to distinguish the general Shannon expression for capacity
(\ref{capacity1}) and particular result for the specific linear additive white
Gaussian noise  channel (\ref{CapacityShannon}). However, recent advances in
fibre-optic communication where the channel is nonlinear, as opposed to the linear
AWGN, changed the situation. To increase the channel capacity over a certain
bandwidth with a  given accumulated noise of optical amplifiers, one has to increase
the signal power, see (\ref{CapacityShannon}). This works in the low SNR limit but
the effect the refraction index's dependence on light intensity dramatically changes
the propagation properties of the channel at higher optical signal power. In other
words, the fibre-optic channel is nonlinear. Recent studies have shown that the
spectral efficiency (that is, the number of bits transmitted per second per Hertz
--- practical characteristics having the same dimension as channel capacity) of a
fibre-optic channel is limited by the Kerr nonlinearity. These studies indicated
that observable spectral efficiency always turns out to be less  than the Shannon
limit of the corresponding linear AWGN channel (\ref{CapacityShannon})
\cite{Mitra:2001,Desurvire:2006,Essiambre:2008,Essiambre:2010,Ellis:2010,Richardson:2010,Killey:2011,N4,Mecozzi:2012}.
It has been observed that  the spectral efficiency of the nonlinear channel
decreases with increasing $\mathrm{SNR}$ at high enough values of $\mathrm{SNR}$
\cite{Mitra:2001,Desurvire:2006,Essiambre:2008,Essiambre:2010,Ellis:2010,N4,Mecozzi:2012}.
This analysis certainly provides only a lower bound on channel capacity and  does
not prove that the Shannon nonlinear fibre channel capacity is decreasing with
power; see, for example,  discussions in
\cite{Agrell:2012,Agrell:2013,Agrell:2014,tdyt03}.

In general, there is a widely spread opinion that nonlinear channel capacity is
always less than the capacity of the corresponding  linear AWGN channel for equal
$\mathrm{SNR}$. However, in Ref. \cite{Turitsyn:2012,Sorokina2014} the authors note
that  nonlinearity can be either destructive or constructive. Moreover, in Ref.
\cite{Agrell:2012} it was proved that the capacity of certain nonlinear channels
could not decrease with $\mathrm{SNR}$. The capacity of nonlinear fibre channels is
still an open problem of great practical and fundamental importance.

In this work, we estimate analytically the first nonzero correction to the mutual
information of the channel described by the NLSE with additive Gaussian noise. We
then calculate in the perturbation theory (at large $\mathrm{SNR}$ and small
nonlinearity) the conditional probability density for the NLSE channel by the method
developed recently in \cite{Terekhov:2014}. Finally, we demonstrate that all
singular in $\mathrm{SNR}$ corrections  to the mutual information are cancelled and
the resulting correction is of order of squared nonlinearity parameter.

The article is organized as follows. In the next section we present the expression
for the conditional probability density. Then in the presented channel model we
calculate the mutual information and classify all possible correction to it at large
$\mathrm{SNR}$ and small nonlinearity parameter. Finally, we present the main
conclusions of the paper. Details of our calculations are placed in the
Supplementary Materials \cite{SupplMat}.

%==================The model of nonlinear channel and conditional probability
\section{The nonlinear channel model and conditional probability}

Let us consider the propagation of the signal  $\psi_\omega(z)$ in  the channel
modelled by the NLSE with AWGN, that we rewrite, for convenience, in the frequency
domain, see \cite{Iannoe:1998,Turitsyn:2000}:
\begin{eqnarray}\label{startingCannelEq}
&&\!\!\!\!\!\!\!\!\!\partial_z \psi_\omega (z)-i\beta\omega^2\psi_\omega (z) -
i\gamma \int \frac{d\omega_1 d\omega_2 d\omega_3}{(2\pi)^3} \times \nonumber\\
&& \!\!\!\!\!\!\!\!\! \times\delta(\omega+\omega_3-\omega_1-\omega_2)
\psi_{\omega_1} (z) \psi_{\omega_2} (z) \bar{\psi}_{\omega_3} (z)= \eta_\omega(z)\,,
\end{eqnarray}
where $\beta$ is the dispersion coefficient, $\gamma$ is the Kerr nonlinearity
coefficient, bar means complex conjugation, $\eta_\omega (z)$ is an additive complex
white noise with zero mean and the correlator $\langle \eta_\omega
(z)\bar{\eta}_{\omega^\prime} (z^\prime)\rangle_{\eta} = 2\pi Q
\delta(z-z^\prime)\delta(\omega-\omega^\prime)\, $, where $Q$ is  the noise power
per unit frequency per unit length \cite{Iannoe:1998,Essiambre:2010}:
$P_{noise}=QLW/2\pi$, here $W/(2 \pi)$ is a frequency bandwidth, and $L$ is the
channel length.

It is worth emphasizing that in a nonlinear channel, transmitted and received signal
bandwidths can be different. Therefore,  we assume here that in general, the input
$X(\omega)$  and output $Y(\omega)$ signals  may have different channel bandwidths
$W$ and $W'$, with $ W' \supset W$.

We introduce the dimensionless parameter $\tilde{\gamma}= P_{ave} \gamma L $ which
describes the impact of nonlinearity. Here the average power of the signal $X$ reads
\begin{align}\label{Pave}
P_{ave}=\lim_{T\to\infty}\int {\cal D}X P[X] \frac{1}{T}\int \frac{d\omega}{2\pi} |X({\omega})|^2\,,
\end{align}
where $T$ is the large time interval containing the whole input signal. It has been
shown in Ref. \cite{Terekhov:2014} that for NLSE channel governed by Eq.
(\ref{startingCannelEq}) in the large SNR limit,
\begin{eqnarray}
\epsilon =1/\mathrm{SNR}=QLW/(2\pi P_{ave})\ll 1,
\end{eqnarray}
the conditional  probability density $P[Y(\omega)| X(\omega)]$ to receive
$\psi_\omega(L)=Y(\omega)$  given the input signal $\psi_\omega(0)=X(\omega)$  can
be written as
\begin{widetext}
\begin{eqnarray}
P[Y(\omega)|X(\omega)]=\exp[-\frac{S[\Psi_{\omega}(z)]}{Q}\Big]
\int^{\tilde{\psi}_{\omega}(L)=0}_{\tilde{\psi}_{\omega}(0)=0} {\cal D} \tilde{\psi}
\exp\Big[
-\frac{1}{Q}\Big\{S[\Psi_{\omega}(z)+\tilde{\psi}_{\omega}(z)]-S[\Psi_{\omega}(z)]\Big\}\Big]\,,\label{PYXquasiclassic}
\\
S[\psi]=\int^L_0 dz \int \frac{d\omega}{2\pi} \Big|\partial_z \psi_{\omega}(z)-i
\beta \omega^2 \psi_{\omega}(z) - i\gamma \int \frac{d\omega_1 d\omega_2
d\omega_3}{(2\pi)^3} \delta(\omega+\omega_3-\omega_1-\omega_2) \psi_{\omega_1}(z)
\psi_{\omega_2}(z) \overline{\psi}_{\omega_3}(z) \Big|^2\,.\label{Action}
\end{eqnarray}
\end{widetext}
Here $\Psi_{\omega}(z)$ is  the so-called ``classical trajectory'' \cite{Feynman} of
the path-integral (\ref{PYXquasiclassic}), that is, the extremum function of the
action (\ref{Action}), i.e. $\delta S[\Psi]=0$~--- see Eq.~(6) in \cite{SupplMat},
with the boundary conditions $\Psi_{\omega}(0)=X(\omega)$,
$\Psi_{\omega}(L)=Y(\omega)$. At small $\tilde{\gamma}$ we can calculate
$P[Y(\omega)|X(\omega)]$, Eq. (\ref{PYXquasiclassic}), analytically using the
perturbation theory developed in Ref. \cite{Terekhov:2014}. There are two types of
terms in the expansion of Eq.~(\ref{PYXquasiclassic}) in $\tilde{\gamma}$. The first
type of perturbative corrections comes from the expansion of exponent
$\exp\left[-{S[\Psi_{\omega}(z)]}/{Q}\right]$ and has the structure
\begin{eqnarray}\label{ExponentExpansion}
&&\exp\left[-\frac{S[\Psi_{\omega}(z)]}{Q}\right]\approx\exp\left[-\frac{S[\Psi^{(0)}_{\omega}(z)]}{Q}\right]_{\gamma=0}\times\nonumber\\
&&\left(1+\sum_{p=0}^\infty\sum_{k=1}^\infty \alpha_{p,k}[\Psi^{(0)}_{\omega}(z)]\tilde{\gamma}^p\left(\frac{\tilde{\gamma}}{\epsilon}\right)^k\right)\,.
\end{eqnarray}
Here $\Psi^{(0)}_{\omega}(z)=e^{i\beta \omega^2 z}\Big[z B(\omega)/L+X(\omega)\Big]$
is the solution at $\gamma=0$ of the equation $\delta S[\Psi]=0$, see  Eqs.~(6) and
(8) in \cite{SupplMat}, with the boundary conditions $\Psi_{\omega}(0)=X(\omega)$,
$\Psi_{\omega}(L)=Y(\omega)$. Here $B(\omega) =e^{-i\beta \omega^2 L}
Y(\omega)-X(\omega) $, and $\alpha_{p,k}[\Psi^{(0)}_{\omega}(z)]$ are some
constructively defined functionals, see explicit expressions in \cite{SupplMat}.
The second type of corrections originates from the expansion of the path-integral in
Eq.~(\ref{PYXquasiclassic}) and has the form
\begin{eqnarray}\label{PathIntegralExpansion}
\Lambda^{(M^\prime)}_{QL}\left(1+\sum_{p=1}^\infty\sum_{k=0}^\infty\gamma_{p,k}
[\Psi^{(0)}_{\omega}(z)]\tilde{\gamma}^p\left(\tilde{\gamma}\epsilon\right)^k\right)\,,
\end{eqnarray}
where $\Lambda^{(M^\prime)}_{QL} = \Big( \frac{\delta}{\pi QL}\Big)^{M^\prime} $ is
the normalization factor and $\gamma_{p,k}[\Psi^{(0)}_{\omega}(z)]$ are some defined
functionals, see \cite{SupplMat}. Since the parameter $\epsilon$ is assumed to be
small, we can use the quasi-classical approach, and the main contribution to
$P[Y|X]$ comes from the expansion of exponent Eq. (\ref{ExponentExpansion}). We
substitute the function $\Psi^{(0)}$ to the right-hand side of Eq.
(\ref{ExponentExpansion}), then the obtained result is multiplied by
(\ref{PathIntegralExpansion}) leading to:
\begin{eqnarray}\label{fullExpansion}
&& P[Y(\omega)|X(\omega)]\approx P^{(0)}[Y(\omega)|X(\omega)]\Biggl(1+\gamma_{1,0}\tilde{\gamma}+\nonumber\\
&&\!\!\!\sum_{k=1}^\infty \alpha_{0,k}\left(\frac{\tilde{\gamma}}{\epsilon}\right)^k
+ \tilde{\gamma}\sum_{k=1}^\infty \left[\alpha_{1,k}+\alpha_{0,k}\gamma_{1,0}\right]
\left(\frac{\tilde{\gamma}}{\epsilon}\right)^k \Biggr)+\nonumber\\
&&{\cal O }\left( \tilde{\gamma}^2\right),
\end{eqnarray}
where
\begin{eqnarray}
 \!\! P^{(0)}[Y(\omega)|X(\omega)]=\Lambda^{(M^\prime)}_{QL}\exp\Big[- \!\frac{1}{QL}\! \!\int_{W'}\!
 \frac{d\omega}{2\pi} |B(\omega)|^2\Big].
\label{P0YX}
\end{eqnarray}
In Eq. (\ref{fullExpansion}) we keep only leading and next-to-leading  order  in
$\tilde{\gamma}$ terms for every order in  $\tilde{\gamma}/\epsilon$. This means
that the parameter $\tilde{\gamma}/\epsilon$ can be of  order of unity. We omit the
dependence of functionals  $\alpha_{p,k}$ and $\gamma_{p,k}$ on
$\Psi^{(0)}_{\omega}(z)$, but have it in mind. Now nonlinear corrections to the
mutual information can be calculated.

%=============Calculation of the nonlinear corrections to the mutual information
\section{ Calculation of the nonlinear corrections to the mutual information}

To calculate the mutual information, we have to calculate the path-integral
(\ref{capacity2}). In the previous section, we derived the expression
(\ref{fullExpansion}) for the conditional probability function.  Let us introduce a
probability function of input signal $P[X(\omega)]$.  In our consideration, the PDF
$P[X(\omega)]$  is chosen to be Gaussian in the spectral domain W and zero in $W'
\setminus W$. In discrete form  $P[X(\omega)]$ reads
\begin{eqnarray}
&\mspace{-20mu}P[X(\omega)]=  \Lambda^{(M)}_{P}\!\! \left(\prod^{M}_{i \in W}
e^{-\frac{\delta}{P} |X_i|^2}\right) \prod^{M'-M}_{j \in W' \setminus W}
\mspace{-2mu} \delta( X_{j}).\,\label{PX}
\end{eqnarray}
where $\delta(X_{j})=\delta( Re\, X_{j}) \delta(Im\, X_{j})$ is  the
$\delta$-function, frequency domain $W$ ($W'$) and is divided by $M$ ($M'$) grid
spacing $\delta = \frac{W}{2\pi M} = \frac{W'}{2\pi M'}$; $X_{j}=X(\omega_j)$. The
coefficient $\Lambda^{(M)}_{P}= \Big( \frac{\delta}{\pi P}\Big)^M $ follows from the
normalization condition $\int {\cal D}X P[X(\omega)] =1$. The measure ${\cal D}X$
 in (\ref{capacity2}), and in (\ref{Pout}), is understood as ${\cal D}X
=\prod_{j=1}^{M'} d Re\, X_j \, d Im \, X_j$.  As follows from the
Nyquist-Shannon-Kotelnikov theorem \cite{Shannon:1949} $M$ should be chosen greater
than $T {W}/{2\pi}$: the limit ${T\to\infty}$ in Eq.~(\ref{Pave}) is equivalent to
$M=T W/2\pi \to \infty$ in the measure. Parameter $P$ in Eq.~(\ref{PX}) describes
the signal power per unit of frequency (power spectral density), so the average
signal power (\ref{Pave}) is $P_{ave}=P W/2\pi \gg P_{noise}$ and the nonlinearity
parameter is $\tilde{\gamma}={\gamma} P L W/(2\pi)$.

First of all, corrections to the mutual information proportional to $\tilde{\gamma}$
vanish. These corrections to the mutual information come only from the term
$\gamma_{1,0}\tilde{\gamma}$ in (\ref{fullExpansion}). From the explicit expression
for $\gamma_{1,0}$  one can see that after integration over fields $X(\omega)$ and
$Y(\omega)$ these contributions vanish as the imaginary part of the real number, see
\cite{SupplMat}. Further, all corrections of order of $(\tilde{\gamma}/\epsilon)^k$
to the mutual information are equal to zero for all $k>0$ as well. We explain this
cancellation in \cite{SupplMat} by the counting of field $B(\omega)=e^{-i\beta
\omega^2 L} Y(\omega)-X(\omega)$ after the change of integration variable from
$Y(\omega)$ to $B(\omega)$ in the path-integral (\ref{capacity2}). Owing to
Eq.~(\ref{P0YX}) the variation scale of $B(\omega)$ in the path-integral is of order
of $\sqrt{QL}$. And every field $B(\omega)$ before the exponent yields the
suppression factor $\sqrt{QL}$ after the integration over $B(\omega)$. The first
next-to-leading in $\tilde{\gamma}$ singular correction  of order of
$\tilde{\gamma}^2/ \epsilon$ is equal to zero after the intricate cancellation of
different terms in the mutual information expression. We present the  details of
this cancellation  in \cite{SupplMat}. Basing on another approach to the mutual
information calculation \cite{TTKhR:2016} we can make a guess that all singular in
$1/\epsilon$ corrections (i.e. of order of
$\tilde{\gamma}^p(\tilde{\gamma}/\epsilon)^k$ for $k>0$ and $p\geq 0$) to the mutual
information should be equal to zero as well. Roughly speaking here we can use the
same arguments  as in the case of the leading singular corrections
$(\tilde{\gamma}/\epsilon)^k$. Finally, after all cancellations we obtain the
following expression for the mutual information:
\begin{eqnarray}
I_{P[X]}=M \log\Big[1+\frac{1}{\epsilon}\Big]+ {\cal O }\left(
\tilde{\gamma}^2\right)+{\cal O }\left( \epsilon \right)\,, \label{MutInfRes}
\end{eqnarray}
where the first term in the right-hand side is the well-known Shannon's  result
(\ref{CapacityShannon}) for the  linear channel \cite{Shannon:1948}, and the second
term is the nonlinear correction to the mutual information. As a coefficient of
${\cal O }\left( \tilde{\gamma}^2\right)$ one has the  function of dimensionless
dispersion parameter $\tilde{\beta}=\beta L W^2$. This function of $\tilde{\beta}$
should be negative at least for small dispersion parameter. Indeed for a
nondispersive nonlinear optical fibre channel one has the exact in nonlinearity
result \cite{TTKhR:2015} with the negative correction:
\begin{eqnarray}
&& \!\!\!\!I^{(\beta=0)}_{P[X]}=M \log\Big[1+\frac{1}{\epsilon}\Big]- \frac{M}{2} \!
\int^{\infty}_{0} \! \!d \tau e^{-\tau} \log\left(1+\frac{\tau^2
\tilde{\gamma}^2}{3} \right)= \nonumber
\\&& \!\!\!\! M \log\Big[1+\frac{1}{\epsilon}\Big]- M\frac{\widetilde{\gamma}^2}{3}+{\cal
O}(\widetilde{\gamma}^{\,4}). \label{MutInfzero}
\end{eqnarray}
Moreover at  large dispersion parameter $\tilde{\beta}$ the function in question
should be decreasing function approaching to the Shannon limit: for large dispersion
parameter one can neglect the nonlinear term in NLSE (\ref{startingCannelEq})
resulting in the linear channel, see Eq.~(\ref{CapacityShannon}). The exact
calculation of the $\tilde{\beta}$-dependence of the coefficient before ${\cal O
}\left( \tilde{\gamma}^2\right)$ in Eq.~(\ref{MutInfRes}) is the matter of our
further considerations \cite{TTKhR:2016}.

%==========================================Conclusion
\section{Conclusion}

We have described the perturbative method to obtain the analytical expression for
the mutual information $I_{P[X]}$ (\ref{capacity2}) of the NLSE channel at large
$\mathrm{SNR}={1}/{\epsilon}$ and small nonlinearity $\tilde{\gamma}\ll 1$. We have
demonstrated that all singular in $\epsilon$ terms in the leading orders
$(\tilde{\gamma}/\epsilon)^k$ and in the first next-to-leading order
$\tilde{\gamma}^2/ \epsilon$ vanish in the expression for the mutual information
$I_{P[X]}$ calculated for the Gaussian PDF $P[X]$. At small nonlinearity
$\tilde{\gamma}$ the first nonlinear correction to the mutual information is of
order of $\tilde{\gamma}^2$, and it is negative at least for small dispersion
parameter $\tilde{\beta}$ and vanishing for large $\tilde{\beta}$. Let us stress
once again that we have considered only the mutual information $I_{P[X]}$ in the
case of a Gaussian input signal PDF $P[X]$ rather than the channel capacity
(\ref{capacity1}). However, the quantity $I_{P[X]}$ is a natural estimate of a low
bound on the channel capacity since it reproduces the Shannon capacity of the linear
AWGN channel (\ref{CapacityShannon}) in the leading order in $\tilde{\gamma}$. The
ultimate calculation of the corrections of order of $\tilde{\gamma}^2$ to the mutual
information will be the matter of our future considerations, see \cite{TTKhR:2016}.

%==========================================Acknowledgments
\emph{\it Acknowledgments.}

The work of I.~S.~Terekhov was supported by the Russian Science  Foundation (RSF)
(grant No. 16-11-10133). Part of the work was supported by the Russian Foundation
for Basic Research (RFBR), Grant No. 16-31-60031/15. A.~V.~Reznichenko thanks the
President program (СП-2415.2015.2) for support. The work of S.~K.~Turitsyn was
supported by the grant of the Ministry  of Education and Science of the Russian
Federation (agreement No. 14.B25.31.0003) and the EPSRC project UNLOC.

%===========================================Bibliography

\end{document}

%\begin{titlepage}
\hskip 2cm
\date{\today}
\title{Supplementary Materials for the article: Mutual information in nonlinear communication
channel. Preliminary analytical results in large SNR and small nonlinearity limit}

\author{I.~S.~Terekhov}
\email[E-mail: ]{I.S.Terekhov@gmail.com}
\affiliation{Budker Institute of  Nuclear Physics of Siberian Branch Russian Academy of Sciences, Novosibirsk, 630090 Russia}
%\affiliation{School of Physics, University of New South Wales, Sydney 2052, Australia}
\affiliation{Novosibirsk State University, Novosibirsk, 630090 Russia}
\author{A.~V.~Reznichenko}
\email[E-mail: ]{A.V.Reznichenko@inp.nsk.su}
\affiliation{Budker Institute of  Nuclear Physics of Siberian Branch Russian Academy of Sciences, Novosibirsk, 630090 Russia}
\affiliation{Novosibirsk State University, Novosibirsk, 630090 Russia}
\author{S.~K.~Turitsyn}
\email[E-mail: ]{s.k.turitsyn@aston.ac.uk} \affiliation{Novosibirsk State
University, Novosibirsk, 630090 Russia} \affiliation{Aston Institute of Photonics
Technologies, Aston University, Aston Triangle, Birmingham, B4 7ET, UK}

\keywords{information entropy, channel capacity, mutual information, nonlinear
Shr\"{o}dinger equation.} \pacs{05.10.Gg, 89.70.-a,  02.70.-c,02.70.Rr,05.90.+m}
\maketitle \maketitle
%==========================================

\section{Conditional probability density function}
In Ref.\cite{Terekhov:2014} we have shown that in the case $1/\epsilon=\mathrm{SNR}
\gg 1$  the conditional probability density function can be written in the form:
\begin{eqnarray}\label{QuasiclassProbabInitial}
P[Y(\omega)|X(\omega)] = e^{-S[\Psi_\omega (z)]/Q} \int\limits_{\tilde{\psi}_\omega
(0)=0}^{\tilde{\psi}_\omega (L)=0}{\cal D}\tilde{\psi}  e^{-(S[\Psi_\omega
(z)+\tilde{\psi}_\omega (z)]-S[\Psi_\omega (z)])/Q}\,,
\end{eqnarray}
where the measure is defined  as
\[
{\cal D} \tilde{\psi}=\lim_{\delta \rightarrow 0}\lim_{\Delta \rightarrow 0} \Big( \frac{\delta}{\Delta \pi Q}\Big)^{N M'}\prod^{M'}_{j=1}\prod^{N-1}_{i=1}dRe\tilde{\psi}_{i,\,j}\,dIm\tilde{\psi}_{i,\,j},
\]
here $\tilde{\psi}_{i,\,j}=\tilde{\psi}_{\omega_j}(z_i)$ and $\Delta=\frac{L}{N}$ is
the  coordinate grid spacing, and  $\delta = \frac{W}{2\pi M} = \frac{W'}{2\pi M'}$
is the frequency grid spacing. The action $S[\psi]$ reads:
\begin{eqnarray}
&&S[\psi]=\int_0^L dz\int\frac{d\omega}{2\pi}\left|{\cal L}[\psi]\right|^2\,,\label{action}\\
&&{\cal L}[\psi]={\cal L}^{(0)}[\psi]-V[\psi]\,,\\
&&{\cal L}^{(0)}[\psi]=\partial_z \psi_\omega (z)-i\beta\omega^2\psi_\omega (z)\,,\\
&&V[\psi]=i\gamma\int
\frac{d\omega_1d\omega_2d\omega_3}{(2\pi)^2}\delta(\omega_1+\omega_2-\omega_3-\omega)
 \psi_{\omega_1} (z) \psi_{\omega_2} (z) \bar{\psi}_{\omega_3} (z)\,.
\end{eqnarray}

The function $\Psi_\omega (z)$, the ``classical trajectory'', is the solution of the
equation $\delta S[\Psi]=0$, see Eq. (14) in \cite{TheArticle:2014}:
\begin{eqnarray}\label{classicalTrajectoryEq}
&&\left(\partial_z-i\beta\omega^2\right)^2\Psi_\omega (z)-\nonumber\\
&&
i\gamma\int\frac{d\omega_1d\omega_2d\omega_3}{(2\pi)^2}\delta(\omega_1+\omega_2-\omega-\omega_3)
\bigg\{4 \Psi_{\omega_2} (z) \bar{\Psi}_{\omega_3} (z)\left[\left(\partial_z-i\beta\omega_1^2\right)\Psi_{\omega_1} (z)\right]-\frac{\mu}{L}\Psi_{\omega_1} (z) \Psi_{\omega_2} (z) \bar{\Psi}_{\omega_3} (z)\bigg\}-\nonumber\\
&&3\gamma^2 \int\frac{d\omega_1d\omega_2d\omega_4 d\omega_5d\omega_6}{(2\pi)^4}
\delta(\omega_1+\omega_2+\omega_4-\omega_5-\omega_6-\omega) \Psi_{\omega_1} (z)
\Psi_{\omega_2} (z) \Psi_{\omega_4} (z)\bar{\Psi}_{\omega_5}
(z)\bar{\Psi}_{\omega_6} (z)=0\,,
\end{eqnarray}
with the boundary conditions: $\Psi_\omega (0)=X(\omega)\,,\quad \Psi_\omega
(L)=Y(\omega)$,  and $\mu=i \beta L (\omega^2+\omega^2_3-\omega^2_1-\omega^2_2)$.
The equation (\ref{classicalTrajectoryEq}) can be solved using perturbation theory
at small $\gamma$. We present the solution $\Psi_\omega(z)$ in the form:
\begin{eqnarray}\label{functionExpansion}
\Psi_\omega(z)=\sum_{k=0}^\infty\Psi_\omega^{(k)}(z)\,,
\end{eqnarray}
where $\Psi_\omega^{(k)}(z)$ is the solution of the Eq.
(\ref{classicalTrajectoryEq}) of order of $\gamma^k$. For calculations of the mutual
information with the precision ${\cal O}(\gamma^2)$ we need only first two terms of
expansion (\ref{functionExpansion}). The solution of
Eq.(\ref{classicalTrajectoryEq}) in the leading and next-to-leading order reads: \be
\Psi^{(0)}_{\omega}(z)=e^{i\beta \omega^2
z}\Big[\frac{z}{L}B(\omega)+X(\omega)\Big], \label{Psi0} \ee here
$B(\omega)=e^{-i\beta \omega^2 L} Y(\omega)-X(\omega)$. The first order correction
reads \be \Psi^{(1)}_{\omega}(z)=i \gamma e^{i\beta \omega^2 z}
\int^{L}_{0}dz'G(z,z') F_{\omega}(z'), \label{Psi1} \ee where
$G(z,z')=\dfrac{z-L}{L}z'+(z'-z)\theta(z'-z)$ is the Green function  of the
$\partial^2_{z}$ operator (with the boundary conditions:
$\Psi^{(1)}_{\omega}(0)=\Psi^{(1)}_{\omega}(L)=0$). In Eq.~(\ref{Psi1}) we have
\begin{equation}
\begin{split}
& F_{\omega}(z)=\int_{W'}
\frac{d\omega_1d\omega_2d\omega_3}{(2\pi)^2}\delta(\omega_1+\omega_2-\omega_3-\omega)
\frac{e^{-\mu z /L}}{L} \Big[\frac{z}{L}B(\omega_2)+X(\omega_2)\Big]\times \\&
\Big[\Big(4-\mu\frac{z}{L}\Big)B(\omega_1)- \mu X(\omega_1) \Big]
\Big[\frac{z}{L}\overline{B}(\omega_3)+\overline{X}(\omega_3)\Big].
\end{split}
\end{equation}

The substitution of the solution $\Psi_\omega(z)$ in the form
(\ref{functionExpansion})  to the action (\ref{action}) results in
\begin{eqnarray}
S[\Psi_\omega(z)]=\sum_{k=0}^\infty S^{(k)}[\Psi_\omega(z)]\,,
\end{eqnarray}
where $ S^{(k)}[\Psi_\omega(z)]$ is the term of order of $\gamma^k$ of the action
(\ref{action}) expansion in $\gamma$. Now we can  expand the exponential prefactor
in Eq. (\ref{QuasiclassProbabInitial}) and obtain functions
$\alpha_{p,k}[\Psi_\omega^{(0)}(z)]$, see Ref. \cite{TheArticle:2014}, Eq.~(9). The expansion
of the exponential factor has the form:
\begin{eqnarray}
\!\!\!\!\!e^{-S[\Psi_\omega
(z)]/Q}=\exp\left\{-\frac{S^{(0)}[\Psi_\omega(z)]}{Q}\right\}\left(1+\sum_{k=1}^\infty\frac{\gamma^k}{k!}\left[\frac{\partial^k}{\partial
\gamma^k}\exp\left\{-\frac{\sum_{p=1}^\infty
S^{(p)}[\Psi_\omega(z)]}{Q}\right\}\right]_{\gamma=0}\right).
\end{eqnarray}
Therefore functions $\alpha_{0,k}[\Psi_\omega^{(0)}(z)]$ and $\alpha_{1,k}[\Psi_\omega^{(0)}(z)]$ have the following form:
\begin{eqnarray}
\alpha_{0,k}[\Psi_\omega^{(0)}(z)]&=&\frac{(-1)^k}{k!}\left(\frac{LW}{2\pi P_{ave}}\frac{S^{(1)}[\Psi_\omega(z)]}{\tilde{\gamma}}\right)^k\,,\label{alpha1}\\
\alpha_{1,k}[\Psi_\omega^{(0)}(z)]&=&
\frac{(-1)^{k}}{(k-1)!(k+1)!}\left(\frac{WL}{2\pi
P_{ave}}\right)^{k}\frac{\partial^{k+1}
\left[S^{(2)}[\Psi_\omega(z)]\left(S^{(1)}[\Psi_\omega(z)]\right)^{k-1}\right]}{\partial
\tilde{\gamma}^{k+1}}\label{alpha2}\,,\quad k\geq1\,,
\end{eqnarray}
where $\tilde{\gamma}={\gamma} P L W/(2\pi)$, and
\begin{eqnarray}
\label{S0} S^{(0)}[\Psi_\omega(z)]&=& \int_0^L dz\int\frac{d\omega}{2\pi}\left|{\cal
L}^{(0)}[\Psi^{(0)}_\omega(z)]\right|^2,
\\
\label{S1} S^{(1)}[\Psi_\omega(z)]&=&2\int\limits_0^L
dz\int_{W'}\frac{d\omega}{2\pi}Re \left\{{\cal
L}^{(0)}[\Psi^{(0)}_\omega(z)]\bar{V}[\Psi^{(0)}_\omega(z)]\right\}=
2\int\limits_0^L \frac{dz}{L}\int_{W'}\frac{d\omega}{2\pi}Re\left\{e^{i\beta\omega^2z} B(\omega)
\bar{V}[\Psi^{(0)}_\omega(z)]\right\}\,,\\
S^{(2)}[\Psi_\omega(z)]&= &\int^{L}_{0}dz \!\int_{W'}\! \frac{d\omega}{2\pi}\Big[
\Big| {\cal L}^{(0)}[\Psi^{(1)}_{\omega}(z)] - \overline{V}[\Psi^{(0)}_{\omega}(z)]
\Big|^2 -2 Re\left\{{\cal
L}^{(0)}[\Psi^{(0)}_\omega(z)]\bar{V}_1[\Psi^{(0)}_\omega(z),\Psi^{(1)}_\omega(z)]\right\}
\Big] \,. \label{S2}
\end{eqnarray}
where ${V}_1[\Psi,\psi]$ is defined below: see Eq. (\ref{V1andV2}). Here in
(\ref{S1}) and (\ref{S2}) we have used that $2 \int dz Re \Big\{ {\cal
L}^{(0)}[\Psi^{(0)}_\omega(z)] \bar{\cal L}^{(0)}[\Psi^{(n)}_\omega(z)]\Big\}=0 $ by
virtue of the boundary conditions:
$\Psi^{(n)}_{\omega}(0)=\Psi^{(n)}_{\omega}(L)=0$.

To calculate the path-integral in Eq. (\ref{QuasiclassProbabInitial}) we  substitute
the solution $\Psi_\omega(z)$ to the action. Here we are interested in only leading
order in $\epsilon$ terms, therefore we keep only quadratic in $\tilde{\psi}_\omega
(z)$ term in the difference $S[\Psi_\omega (z)+\tilde{\psi}_\omega
(z)]-S[\Psi_\omega (z)]$ since terms with higher order in $\tilde{\psi}_\omega (z)$
are suppressed in the parameter $\epsilon$. In the leading order in $\epsilon$ the
path-integral can be written in the form:
\begin{eqnarray}\label{PathIntegral}
&&\int\limits_{\tilde{\psi}_\omega (0)=0}^{\tilde{\psi}_\omega (L)=0}{\cal D}\tilde{\psi} e^{-(S[\Psi_\omega (z)+\tilde{\psi}_\omega (z)]-S[\Psi_\omega (z)])/Q}\approx\nonumber\\
&&\int\limits_{\tilde{\psi}_\omega (0)=0}^{\tilde{\psi}_\omega (L)=0}{\cal D}\tilde{\psi}\exp\left\{-\frac{1}{Q}\int_0^L dz\int\frac{d\omega}{2\pi}\left|{\cal L}^{(0)}[\tilde{\psi}]\right|^2\right\}\exp\left\{-\frac{\Delta S[\Psi,\tilde{\psi}]}{Q}\right\}\,,
\end{eqnarray}
where
\begin{eqnarray}\label{DeltaS}
\Delta S[\Psi,\tilde{\psi}]&=&\int_0^L\! dz\!\!\int\frac{d\omega}{2\pi}\Bigg[\left|V_1[\Psi,\tilde{\psi}]+V_2[\Psi,\tilde{\psi}]+V[\tilde{\psi}]\right|^2-\nonumber\\
&&2Re\left\{{\cal L}^{(0)}[\tilde{\psi}]\bar{V}_1[\Psi,\tilde{\psi}]+\left({\cal L}[\Psi]+{\cal L}^{(0)}[\tilde{\psi}]\right)\left(\bar{V}_2[\Psi,\tilde{\psi}]+\bar{V}[\tilde{\psi}]\right)\right\}\Bigg]\,.
\end{eqnarray}
When deriving (\ref{DeltaS}) we have used the equation of motion (\ref{classicalTrajectoryEq}).
\begin{eqnarray} \label{V1andV2}
V_1[\Psi,\tilde{\psi}]&=&i\gamma \int\frac{d\omega_1 d\omega_2 d\omega_3}{(2\pi)^2}\delta(\omega_1 +\omega_2- \omega_3 -\omega) \left( 2 \tilde{\psi}_{\omega_1}(z) \Psi_{\omega_2}(z)\bar{\Psi}_{\omega_3}(z)+\bar{\tilde{\psi}}_{\omega_3}(z) \Psi_{\omega_1}(z)\Psi_{\omega_2}(z)\right)\,,\\
V_2[\Psi,\tilde{\psi}]&=&i\gamma \int\frac{d\omega_1 d\omega_2
d\omega_3}{(2\pi)^2}\delta(\omega_1 +\omega_2- \omega_3 -\omega) \left( 2
\tilde{\psi}_{\omega_1}
(z)\Psi_{\omega_2}(z)\bar{\tilde{\psi}}_{\omega_3}(z)+\tilde{\psi}_{\omega_1}(z)
\tilde{\psi}_{\omega_2}(z)\bar{\Psi}_{\omega_3}(z)\right)\,.
\end{eqnarray}
The functional $\gamma_{p,0}$ has the form
\begin{eqnarray}\label{gamma}
\gamma_{p,0}[\Psi_\omega^{(0)}(z)]=\frac{1}{P^{(0)}[0,0]}\int\limits_{\tilde{\psi}_\omega (0)=0}^{\tilde{\psi}_\omega (L)=0}{\cal D}\tilde{\psi}\exp\left\{-\frac{1}{Q}\int_0^L dz\int\frac{d\omega}{2\pi}\left|{\cal L}^{(0)}[\tilde{\psi}]\right|^2\right\}\frac{1}{p!}\left(\frac{\partial^p }{\partial \tilde{\gamma}^p}\exp\left\{-\frac{\Delta \tilde{S}[\Psi,\tilde{\psi}]}{Q}\right\}\right)_{\gamma=0}\,,
\end{eqnarray}
where
\begin{eqnarray}\label{DeltaStilde}
\Delta \tilde{S}[\Psi,\tilde{\psi}]=\int_0^L\! dz\!\!\int\frac{d\omega}{2\pi}\Bigg[\left|V_1[\Psi,\tilde{\psi}]\right|^2-2Re\left\{{\cal L}^{(0)}[\tilde{\psi}]\bar{V}_1[\Psi,\tilde{\psi}]+{\cal L}[\Psi]\bar{V}_2[\Psi,\tilde{\psi}]\right\}\Bigg]\,.
\end{eqnarray}
The function $P^{(0)}[Y|X]$ has the form, see Ref.\cite{Terekhov:2014}:
\begin{eqnarray} P^{(0)}[Y|X]=\Lambda^{(M')}_{QL}\exp\Big[-\! \frac{1}{QL}\!
\int_{W'}\! \frac{d\omega}{2\pi} |B(\omega)|^2\Big]\,, \label{P0YX}
\end{eqnarray}
\begin{eqnarray}
\Lambda^{(M')}_{D}= \Big( \frac{\delta}{\pi D}\Big)^{M'}\,. \label{Lambda}
\end{eqnarray}

We need only $\gamma_{1,0}$ and $\gamma_{2,0}$ for our calculation. To calculate
these functionals we use method developed in  \cite{Terekhov:2014}. Direct
calculation of $\gamma_{1,0}$  gives:
\begin{eqnarray}\label{gamma10}
\gamma_{1,0}\left[\Psi^{(0)}(z)\right]=\frac{2 W'}{\pi LP_{ave}} Im\,\Bigg\{
\int^{L}_{0} dz \frac{z(L-z)}{L} \int_{W'} \frac{d\omega}{2\pi} {\cal
L}^{(0)}\left[\Psi_{\omega}^{(0)}(z)\right]{\bar{\Psi}_{\omega}^{(0)}(z)}\Bigg\}\,.
\end{eqnarray}

%\begin{eqnarray}
%&&\gamma_{2,0}\left[\Psi^{(0)}(z)\right]=\frac{2 W'}{\pi ( LP_{ave})^2}
%Im\,\Bigg\{  \int^{L}_{0} dz \frac{z(L-z)}{L} \int^{W'/2}_{-W'/2} \frac{d\omega}{2\pi}\left[
%{\cal L}^{(0)}\left[ \Psi_{\omega}^{(1)}(z)\right]{\bar{\Psi}_{\omega}^{(0)}(z)}+
%{\cal
%L}^{(0)}\left[\Psi_{\omega}^{(0)}(z)\right]{\bar{\Psi}_{\omega}^{(1)}(z)}\right]\Bigg\}-
%\nonumber\\
%&&\frac{W'}{(LP_{ave})^2}   \int^{L}_{0} dz \frac{z(L-z)}{L} \int^{W'/2}_{-W'/2}\Bigg[
%5\frac{d\omega_1 d\omega_2 d\omega_3 d\omega_4}{(2\pi)^4} \delta(\omega_1 +\omega_2-
%\omega_3 -\omega_4)
%\Psi_{\omega_1}^{(0)}(z)\Psi_{\omega_2}^{(0)}(z)\bar{\Psi}_{\omega_3}^{(0)}(z)\bar{\Psi}_{\omega_4}^{(0)}(z) +\nonumber\\
%&&
%4 \int^{W'/2}_{-W'/2} \frac{d \omega_2 d \omega_3 d \omega_4  }{(2\pi)^3} \int^{W'+\omega_3}_{-W'+\omega_3} \frac{d \omega_1}{(2\pi)}\, \delta(\omega_1 +\omega_2-
%\omega_3 -\omega_4)
%\Psi_{\omega_1}^{(0)}(z)\Psi_{\omega_2}^{(0)}(z)\bar{\Psi}_{\omega_3}^{(0)}(z)\bar{\Psi}_{\omega_4}^{(0)}(z) \Pi \Big(\frac{\omega_1-\omega_3}{W'}\Big) \Bigg]\,,
%\end{eqnarray}
%where $\Pi(\Omega)=(1-\Omega)\theta(1-\Omega)\theta(\Omega)+(1+\Omega)\theta(1+\Omega)\theta(-\Omega)$.

Using Eqs. (\ref{alpha1})--(\ref{alpha2}) and (\ref{gamma}) we can write
conditional probability density function with accuracy $\tilde{\gamma}^2$ in the
form:
\begin{eqnarray}\label{fullExpansion}
P[Y(\omega)|X(\omega)]\approx P^{(0)}[Y(\omega)|X(\omega)]\Biggl(1 + \alpha^{(1)}[Y|X]+ \alpha^{(2)}[Y|X]\Biggr)\,,
\end{eqnarray}
where we introduce following notations:
\begin{eqnarray} \label{alpna1explicit}
\alpha^{(1)}[Y|X]&=&\alpha_{0,1}\left[\Psi^{(0)}(z)\right]\frac{\tilde{\gamma}}{\epsilon}+
\gamma_{1,0}\left[\Psi^{(0)}(z)\right]\tilde{\gamma}\,,\\
\label{alpna2explicit}
\alpha^{(2)}[Y|X]&=&\alpha_{0,2}\left[\Psi^{(0)}(z)\right]\frac{\tilde{\gamma}^2}{\epsilon^2}
+
\left\{\alpha_{1,1}\left[\Psi^{(0)}(z)\right]+\alpha_{0,1}\left[\Psi^{(0)}(z)\right]
\gamma_{1,0}\left[\Psi^{(0)}(z)\right]\right\} \frac{\tilde{\gamma}^2}{\epsilon} +
\gamma_{2,0}\left[\Psi^{(0)}(z)\right] \tilde{\gamma}^2.
\end{eqnarray}
We note that from Eq.~(\ref{alpha1}) one has
\begin{eqnarray}\label{alpha02}
\alpha_{0,2}\left[\Psi^{(0)}(z)\right]=
{\Big(\alpha_{0,1}\left[\Psi^{(0)}(z)\right]\Big)^2}/{2}.
\end{eqnarray}

Now we are ready to calculate the  mutual information, see Eq. (2) of
\cite{TheArticle:2014}, with accuracy ${\cal O}(\tilde{\gamma}^2)$.

\section{Calculation of the Mutual Information}
To calculate the  mutual information
\begin{eqnarray}
\begin{split} % I deleted \lim_{T\to\infty}\frac{1}{T}
&I_{P[X]}= \int {\cal D}X {\cal D}Y   P[X] P[Y| X]  \log\Big[ \frac{P[Y|
X]}{P_{out}[Y]}\Big], \label{capacity2}
\end{split}
\end{eqnarray}
let us first present the auxiliary correlation functions:
\begin{eqnarray}
\langle B(\omega) \overline{B}(\omega') \rangle_{P^{(0)}[Y|X]}=\int{\cal D}Y
P^{(0)}[Y|X]B(\omega) \bar{B}(\omega') = 2\pi QL\delta(\omega-\omega'),
\label{Bpairing}
\end{eqnarray}
\begin{eqnarray} \langle X(\omega) \overline{X}(\omega')
\rangle_{P[X]}=\int{\cal D}X P[X]X(\omega) \overline{X}(\omega')=2\pi
P\delta(\omega-\omega') \chi_{W}(\omega) \chi_{W}(\omega'), \label{Xpairing}
\end{eqnarray}
where we've used as the integration weights PDF $P^{(0)}[Y|X]$, see (\ref{P0YX}),
and the input signal PDF $P[X]$ of the form
\begin{eqnarray}
&\mspace{-20mu}P[X(\omega)]=  \Lambda^{(M)}_{P}\!\! \left(\prod^{M}_{i \in W}
e^{-\frac{\delta}{P} |X_i|^2}\right) \prod^{M'-M}_{j \in W' \setminus W}
\mspace{-2mu} \delta( X_{j}).\,\label{PX}
\end{eqnarray}
see  Ref.\cite{TheArticle:2014}. Here  $\delta(X_{j})=\delta( Re\, X_{j})
\delta(Im\, X_{j})$ is the $\delta$-function, frequency domain $W$ ($W'$) is divided
by $M$ ($M'$) grid spacing $\delta = \frac{W}{2\pi M} = \frac{W'}{2\pi M'}$;
$X_{j}=X(\omega_j)$. The measure ${\cal D}X$ reads ${\cal D}X=\prod_{j=1}^{M'}d Re
X_j d Im X_j$, and
$\chi_{W}(\omega)=\theta(\frac{W}{2}-\omega)\theta(\frac{W}{2}+\omega)$ stands for
the indicator of the (cyclic) frequency domain $W$ in Eq.~(\ref{Xpairing}).
Equations (\ref{Bpairing}) and (\ref{Xpairing}) can be easily obtained using
discrete form of path-integrals. Since functions $P^{(0)}[Y|X]=P^{(0)}[B]$ and
$P[X]$ have Gaussian form, we can use the Wick theorem \cite{IZ}:
\begin{eqnarray}\label{averB}
\langle B(\omega_1) \rangle_{P^{(0)}[B]}=0\,,
\end{eqnarray}
\begin{eqnarray}\label{averB2}
\langle B(\omega_1)B(\omega_2)\overline{B}(\omega_3)\overline{B}(\omega_4) \rangle_{P^{(0)}[B]}&=&\langle B(\omega_1)\overline{B}(\omega_3)\rangle_{P^{(0)}[B]}\langle B(\omega_2)\overline{B}(\omega_4) \rangle_{P^{(0)}[B]}+\nonumber\\
&&\langle B(\omega_1)\overline{B}(\omega_4) \rangle_{P^{(0)}[B]}\langle B(\omega_2)\overline{B}(\omega_3)\rangle_{P^{(0)}[B]}\,.
\end{eqnarray}
Averaging $\langle...\rangle_{P[X]}$ has the same properties. Using these properties
and Eqs. (\ref{Bpairing}), (\ref{Xpairing}), the calculation of the mutual
information with accuracy ${\cal O}(\tilde{\gamma}^2)$ turns to the simple
calculation of the correlation functions. Note, that now $P_{ave}=P W/(2\pi) \gg
P_{noise}=Q L W/(2\pi)$, nonlinearity parameter $\tilde{\gamma}={\gamma}L
PW/(2\pi)$, and $\mathrm{SNR}={1}/{\epsilon}={P}/{(QL)}$.

%==================================
\subsection{Cancellation of the leading corrections in $1/\epsilon$}

Let us show that all leading corrections in $1/\epsilon$, i.e. the corrections of
order of $(\tilde{\gamma}/\epsilon)^k$, to the mutual information are equal to zero
for all $k>0$. To show that, we substitute expression (\ref{fullExpansion})  to Eq.
(\ref{capacity2}) and change the integration variables from $Y(\omega)$ to
$B(\omega)$. The function $P^{(0)}[Y(\omega)|X(\omega)]$, see Eq.~ (\ref{P0YX}),
depends only on $B(\omega)$ and has a variation scale of order $\sqrt{QL}$, see Eq.~
(\ref{Bpairing}), the scale of variation of function $P[X(\omega)]$ is $P$, which
obeys the condition $P\gg QL$, and the correction of order of
$(\tilde{\gamma}/\epsilon)^k$ comes from the coefficients $\alpha_{0,k}$, but these
coefficients $\alpha_{0,k}$ are explicitly proportional to $B^k(\omega)$, see Eqs.
(\ref{alpha1}), (\ref{S1}). In what follows all terms with odd powers $B(\omega)$
give zero after integration over $B(\omega)$, while terms proportional to even
powers $B^k(\omega)$ (with equal number of the fields $B$ and conjugated fields
$\overline{B}$), after integration over $B(\omega)$, contribute proportionally to
factor $(QL)^{k} \propto \epsilon^k$: additional conjugated fields $\overline{B}$
come from $\bar{V}[\Psi^{(0)}_\omega(z)]$, see Eq.~(\ref{S1}). It reduces the power
of $\epsilon$ in the denominator. Therefore, all corrections to the mutual
information having the form of a power of $\tilde{\gamma}/\epsilon$ disappear.

%==================================
\subsection{Cancellation of the corrections of order of $\tilde{\gamma}$}

It is easy to see that all corrections to the mutual information of order of
$\tilde{\gamma}$ are equal to zero.  These corrections come from the term
$\gamma_{1,0}\tilde{\gamma}$ in Eq.~(\ref{alpna1explicit}). Using the explicit
expression (\ref{gamma10}) one can see that this contribution vanishes after the
integration over  $B(\omega)$ either as the imaginary part of the real number or as
the odd power of $B(\omega)$.

%==================================
\subsection{Cancellation of the sub-leading corrections of order of $\tilde{\gamma}^2/\epsilon$}

To calculate the corrections connected with channel nonlinearity we substitute the
expression (\ref{fullExpansion}) to  Eq. (\ref{capacity2}) and obtain the following
expansion with accuracy $\tilde{\gamma}^2$ for the mutual information:
\begin{eqnarray}\label{MutInfExp}
I_{P[X]}\approx \int\!\! {\cal D} X{\cal D} Y P[X] P^{(0)}[Y| X]\left\{ \log\left[
\frac{P^{(0)}[Y| X]}{P^{(0)}_{out}[Y]}\right]\left(1+\alpha^{(1)}[Y|X]+
\alpha^{(2)}[Y|X]\right)
+\frac{(\alpha^{(1)}[Y|X])^2-(\beta^{(1)}[Y])^2}{2}\right\},
\end{eqnarray}
where $\alpha^{(1)}[Y|X])$ is defined in Eq.~(\ref{alpna1explicit}), and \be
\begin{split}
P^{(0)}_{out}[Y]=\int {\cal D} X P[X] P^{(0)}[Y| X],
\end{split}
\ee
\be
\begin{split}
&\beta^{(1,2)}[Y(\omega)] =\frac{\int {\cal D} X P[X]
P^{(0)}[Y|X]\alpha^{(1,2)}[Y|X]}{P^{(0)}_{out}[Y]}. \label{beta}
\end{split}
\ee The direct calculation of $P^{(0)}_{out}[Y]$ and $\beta^{(1)}[Y(\omega)]$ gives
\begin{eqnarray} P^{(0)}_{out}[Y]= \Lambda^{(M'-M)}_{QL}\exp\Big\{-\frac{1}{QL}\!\!\!\int_{W'
\setminus W}\!\!\!\!\!\frac{d\omega }{2\pi}|Y(\omega)|^2 \Big\}\,
\Lambda^{(M)}_{P+QL} \exp\Big\{-\frac{1}{P+QL}\int_{ W}\!\frac{d\omega
}{2\pi}|Y(\omega)|^2 \Big\}. \label{Poutzero}
\end{eqnarray}

%============new, see#5.2 verso
\begin{eqnarray}
&& \beta^{(1)}[Y]=\frac{ 2 \gamma P}{QL (P+QL)} \int^{L}_{0}dz \int_{W'}
\frac{d\omega d\omega_1 d\omega_2 d\omega_3}{(2\pi)^3}
\delta(\omega_1+\omega_2-\omega_3-\omega)\frac{e^{\mu (z /L-1)}}{2 i} {Y}({\omega} )
{Y}({\omega_3} )\overline{Y}({\omega_1} )\overline{Y}({\omega_2} )\times \nonumber
\\&& \left\{ \chi_{W'\setminus W}(\omega)\chi_{W}(\omega_1)-
\chi_{W'\setminus W}(\omega_1)\chi_{W}(\omega) \right\} \left[ \chi_{W'\setminus
W}(\omega_2)\frac{z}{L}+\chi_{W}(\omega_2)\left(1+\frac{QL}{P+QL}(z /L-1)\right)
\right]\times \nonumber
\\&& \left[ \chi_{W'\setminus
W}(\omega_3)\frac{z}{L}+\chi_{W}(\omega_3)\left(1+\frac{QL}{P+QL}(z /L-1)\right)
\right],\label{beta1}
\end{eqnarray}
where $\mu=i \beta L (\omega^2+\omega^2_3-\omega^2_1-\omega^2_2)$,
$\chi_{W}(\omega)=\theta(\frac{W}{2}-\omega)\theta(\frac{W}{2}+\omega)$, and
$\chi_{W'\setminus W}(\omega)=\chi_{W'}(\omega)-\chi_{W}(\omega)$. Note that if
$W'=W$  then $\beta^{(1)}[Y]$ vanishes. We remind that $W'$ is auxilary bandwidth
containing the bandwidth $W$ where signal $X(\omega)$ is not zero, see Eq.
(\ref{PX}).

The output PDF (\ref{Poutzero}) results in the following pairing on the analogy of
Eqs. (\ref{Bpairing}), (\ref{Xpairing}):
\begin{eqnarray}
\langle Y(\omega) \overline{Y}(\omega') \rangle_{P^{(0)}_{out}[Y]}=\int{\cal D}Y
P^{(0)}_{out}[Y] Y(\omega) \overline{Y}(\omega')=2\pi \delta(\omega-\omega')\left\{
QL\chi_{W'\setminus W}(\omega)  +(P+QL)\chi_{W}(\omega)\right\}. \label{Ypairing}
\end{eqnarray}

We present the expression (\ref{MutInfExp}) for the mutual information in the form:
\begin{eqnarray}
I_{P[X]}=I_0+I_1+I_2+I_3\,,
\end{eqnarray}
here
\begin{eqnarray} I_0= M \log\Big[1+\frac{P}{QL}\Big] \int {\cal D} X {\cal D} Y
P[X] P^{(0)}[Y|X] \Big\{1+\alpha^{(1)}[Y|X]+ \alpha^{(2)}[Y|X] \Big\}\,, \label{I0}
\end{eqnarray}
where the logarithm $\log\left[1+\frac{P}{QL}\right]$  occurs from the normalization
factors (\ref{Lambda}) in Eqs.~(\ref{P0YX}) and (\ref{Poutzero}).

\begin{eqnarray}
I_1= \frac{1}{2}  \int {\cal D} X {\cal D} Y P[X]
P^{(0)}[Y|X]\Big( \alpha^{(1)}[Y|X]\Big)^2,
\label{I1} \end{eqnarray}

\begin{eqnarray}% испр.
&& I_2= \int {\cal D} X {\cal D} Y P[X] P^{(0)}[Y|X]\int_{W'}
\frac{d\omega}{2\pi}\Big(\frac{\chi_{W}(\omega)}{P+QL}|Y(\omega)|^2+\frac{\chi_{W'\setminus
W}(\omega)}{QL}|Y(\omega)|^2-\frac{|B(\omega)|^2}{QL}\Big)\times\\&&
\Big\{1+\alpha^{(1)}[Y|X]+ \alpha^{(2)}[Y|X] \Big\} , \label{I2}
\end{eqnarray}

\begin{eqnarray}
I_3=-\frac{1}{2} \int {\cal D} Y
P^{(0)}_{out}[Y] \Big(\beta^{(1)}[Y]\Big)^2. \label{I3}
\end{eqnarray}

Here the terms (\ref{I0}) and (\ref{I2}) come from the term in (\ref{MutInfExp})
which is proportional to $\log\left[P^{(0)}[Y| X]/P^{(0)}_{out}[Y]\right]$.

For the simplicity of the result presentation we assume that  $W'=W$, and therefore
$\beta^{(1)}[Y]$, see Eq. (\ref{beta1}), vanishes. This assumption kills the
contribution $I_3$, see Eq.~(\ref{I3}).

The direct calculation of $I_0$ shows that path-integral \be \int {\cal D} X {\cal
D} Y P[X] P^{(0)}[Y|X] \Big\{\alpha^{(1)}[Y|X]+ \alpha^{(2)}[Y|X] \Big\}=0\,. \ee It
is the consequence of the normalization condition ($\int{\cal D}YP[Y| X]=1$) for the
conditional probability function $P[Y|X]$:
\begin{eqnarray}\label{normalizationconadition}
\int{\cal D}Y P[Y| X]=1\,
\end{eqnarray}
and therefore
\begin{eqnarray}\label{normalizationPYX2}
\int {\cal D}Y P^{(0)}[Y(\omega)|X(\omega)]\alpha^{(1,2)}[Y|X]=0\,.
\end{eqnarray}

Thus one has the result \be I_0=C_{SH} =M \log\Big[1+\frac{P}{QL}\Big]\,,
\label{I0Res} \ee which coincides with the classical result for the linear channel
capacity $C_{SH}$ (Shannon-Hartley theorem \cite{Shannon:1948}).

Now we proceed to calculate the singular terms (of order of
$\tilde{\gamma}^2/\epsilon$) in the contribution $I_1$, see Eq.~(\ref{I1}). In the
beginning we change the  integration variable from $Y(\omega)$ to
$B(\omega)=e^{-i\beta \omega^2 L} Y(\omega)-X(\omega)$ in the path-integral
~(\ref{I1}). Every pairing (\ref{Bpairing}) results in extra $Q$-suppressed factor.
That is why the most singular term of order of $\tilde{\gamma}^2/\epsilon$ emerges
from one pairing (\ref{Bpairing}) when we retain  only the the first term (singular
in $\epsilon$) in (\ref{alpna1explicit}) and take only the linear in $B(\omega)$
part of this term. After the pairing (\ref{Bpairing}) we perform the straightforward
integration over $X$ in the path-integral ~(\ref{I1}) using the correlator
(\ref{Xpairing}) and Wick theorem. Finally, we present the result of the calculation
of the term $I_1$ in the form:
\begin{eqnarray}
I_1=4M G(\tilde{\beta})\frac{\tilde{\gamma}^2}{\epsilon} \, +{\cal
O}(\tilde{\gamma}^2), \label{Calpha1}
\end{eqnarray}
where the function $G(\tilde{\beta})$ of dimensionless dispersion parameter
$\tilde{\beta}=\beta L W^2$ is defined as
\begin{eqnarray}
\!G(\tilde{\beta})=1+\frac{1}{2{\tilde{\beta}}^2} \int_{\Omega} dy dy_1 dy_2
\frac{\sin^2[\tilde{\beta}(y-y_1)(y-y_2)]}{(y-y_1)^2(y-y_2)^2},
\end{eqnarray}
with $\Omega=[-\frac{1}{2}, \frac{1}{2}]\times [-\frac{1}{2}, \frac{1}{2}]\times
[-\frac{1}{2}, \frac{1}{2}]$ being a simple cubic region.

The next contribution to  $I_{P[X]}$ is $I_2$. When following the singular in
$\epsilon$ terms we can neglect the first term in the parenthesis in Eq.~(\ref{I2}).
We omit the second term in the parenthesis in Eq.~(\ref{I2}) as well since in the
current consideration we assume $W'=W$. Now we change the  integration variables
from $Y(\omega)$ to $B(\omega)$. And for the third term in the parenthesis there are
two possible pairing of the type (\ref{Bpairing}).  The first $Q$-unsuppressed term
(i.e. the term of order of ${\tilde{\gamma}^2}/{\epsilon}$) originates from the
inner pairing (i.e. integration over $B$) of $|B(\omega)|^2$-term:
\[
\frac{1}{QL}\int_{W'} \frac{d\omega}{2\pi}\langle |B(\omega)|^2 \rangle_{P^{(0)}[Y|X]}=M'.
\]
However, for this case $\alpha^{(1,2)}[Y|X]$ contributions from the brace in
Eq.~(\ref{I2}) vanish after integration over $X$ owing to the normalization
condition (\ref{normalizationPYX2}). The second $Q$-unsuppressed term comes from two
pairings of the field $B(\omega)$ from the parenthesis with the first part of
$\alpha^{(2)}[Y|X]$ in the brace, see the first term (of order of
${\tilde{\gamma}^2}/{\epsilon^2}$) in Eq. ~(\ref{alpna2explicit}). These pairings,
see Eq.~(\ref{alpha02}) and (\ref{I1}), result in the same contributions as in
Eq.~(\ref{Calpha1}) but with the opposite sign in the singular term
${\tilde{\gamma}^2}/{\epsilon}$. And one has with accuracy ${\cal
O}(\tilde{\gamma}^2)$:
\begin{eqnarray} I_2=-4M
G(\tilde{\beta})\frac{\tilde{\gamma}^2}{\epsilon} \, +{\cal O}(\tilde{\gamma}^2).
\label{Calpha2}
\end{eqnarray}

The last component $I_3$ of  $I_{P[X]}$ is zero (in considered case $W'=W$):
\begin{eqnarray}
I_3=-\frac{1}{2} \int {\cal D} Y P^{(0)}_{out}[Y] \Big(\beta^{(1)}[Y]\Big)^2= 0.
\end{eqnarray}

Finally, all singular terms of order of ${\tilde{\gamma}^2}/{\epsilon}$ are
cancelled, and in the expression for the spectral efficiency ($i_{P[X]}$)  we obtain
Shannon's logarithm with the corrections ${\cal O }\left( \tilde{\gamma}^2\right)$:
\begin{eqnarray}
&&i_{P[X]}=\frac{I_{P[X]}}{M}\approx\log\Big[1+\frac{1}{\epsilon}\Big]+
{\cal O }\left( \tilde{\gamma}^2\right) + {\cal O }\left(\epsilon\right) =\nonumber\\
&&\!\!\!\log\Big[1+\mathrm{SNR}\Big]+ \mathrm{SNR}^2 \times {\cal O }\left( \gamma^2
L^2 P^2_{noise}\right)\,+ {\cal O} (1/\mathrm{SNR}). \label{finalresult}
\end{eqnarray}

Note that the cancellation of the terms of order of $\tilde{\gamma}^2/ \epsilon$ and
our result (\ref{finalresult}) hold true in general case when $W' \supset W$.
However in this case the cancellation is much more intricate. For example, in this
case the term $I_3$, see Eq.~(\ref{I3}), does contribute in the order
$\tilde{\gamma}^2/ \epsilon$. The cancellation takes place within the expression for
the output signal entropy $H[Y]$ (we remind that $I_{P[X]}=H[Y]-H[Y|X]$) that can be
written with $\gamma^2$ accuracy in the form
\begin{align}\label{HY2}
& H[Y]= M-\log\Lambda^{(M)}_{P+QL}+(M'-M)-\log\Lambda^{(M'-M)}_{QL}-\nonumber \\&
\int{\cal D}Y P^{(0)}_{out}[Y]\Big\{\left( \beta^{(1)}[Y]+\beta^{(2)}[Y]
\right)\Big[-\frac{1}{QL}\int_{W' \setminus W}\!\!\! \frac{d\omega}{2\pi}
|Y(\omega)|^2-\frac{1}{P+QL}\int_{W} \frac{d\omega}{2\pi} |Y(\omega)|^2
\Big]+\frac{\beta^{(1)}[Y]^2}{2}\Big\}.
\end{align}
We checked directly the cancellation of the contributions of order of
$\tilde{\gamma}^2/ \epsilon$ in the mutual information in the case when $W' \supset
W$ by the computer algebra methods using the Wick theorem and the correlators
(\ref{Xpairing}), (\ref{Bpairing}), and (\ref{Ypairing}).

In our further considerations we will calculate the first nonvanishing  corrections
to the mutual information of order of $\tilde{\gamma}^2$, see \cite{TTKhR:2016}.

%===========================================Bibliography